\documentclass[a4paper,10pt]{article}

\def\eq#1{{Eq.~(\ref{#1})}}

\def\pdsl#1#2{(\partial#1/\partial#2)}        
           
\def\pdlr#1#2{\left(\frac{\partial#1}{ \partial#2}\right)}
\def\g{\sqrt{-g}\,}
\def\mr{\mathcal{R}}
\def\rc#1#2#3#4{{R^{#1}_{\phantom{#1}#2#3#4}}}
\def\pc#1#2#3#4{{P_{#1}^{\phantom{#1}#2#3#4}}}
\newcommand{\LL}{Lanczos-Lovelock }
\def\gu#1#2{{g^{#1#2}}}

\title{Some aspects of field equations in generalized theories of gravity}
\author{T. Padmanabhan\\
IUCAA, Pune University Campus,\\
Pune, India.}

\date{ }

\begin{document}

\maketitle

\begin{abstract}
A class of theories of gravity based on a Lagrangian $L = L(\rc abcd, g^{ab})$ which depends on the curvature and metric --- but not on the derivatives of the curvature tensor --- is of interest in several contexts including in the development of the paradigm that treats gravity as an  emergent phenomenon. This class of models contains, as an important subset, all \LL\ models of gravity. I derive several  identities and properties which are useful in the study of these models and clarify some of the issues  that seem to have received insufficient attention in the past literature.
\end{abstract}

\section{Introduction}

The simplest geometrical theory of gravity is Einstein's theory for which the gravitational Lagrangian is proportional to the Ricci scalar. A more general class of theories can be obtained from Lagrangians which have an arbitrary dependence on the curvature tensor and the metric but do not depend on the covariant derivatives of the curvature tensor. In recent years there is a growing recognition that the field equations of gravity may be emergent and may have the same conceptual status as the equations of fluid mechanics of elasticity (for a recent review, see \cite{tprop}). In particular, the \LL\ models of gravity [which is a subset of theories built from Lagrangians of the form $L(\rc abcd, g^{ab})$] exhibit thermodynamic properties  that are strikingly similar to those in Einstein's theory of gravity. The  main purpose of this paper is to describe several curious and useful identities and mathematical results which arise in this general class of theories. To the extent I know, they do not seem to have been emphasized in a coherent manner in the existing literature.

\subsection{Motivation}

Consider a generally covariant metric theory of gravity in $D$ dimensional spacetime based on the action functional $A_{\rm tot} = A_g + A_{\rm matter}$ where the gravitational action is given by 
\begin{equation}
A_g= \int_\mathcal{V} d^Dx\sqrt{-g}\, L[g^{ab},R^a_{\phantom{a}bcd}] 
\label{genaction1}
\end{equation} 
The variation of this action under the variation $g^{ab} \to g^{ab} + \delta g^{ab}$ can be computed in a straightforward manner \cite{TPtext} to give the result:
\begin{equation}
\delta A_g = \delta \int_\mathcal{V} d^Dx\sqrt{-g}\, L = \int_\mathcal{V} d^Dx \, \sqrt{-g} \, E_{ab} \delta g^{ab} + 
\int_\mathcal{V} d^Dx \, \sqrt{-g} \, \nabla_j \delta v^j
\label{tpgrav2}
\end{equation}
where
\begin{equation}
\sqrt{-g}E_{ab}\equiv\left( \frac{\partial \sqrt{-g}L}{\partial g^{ab}} 
-2\sqrt{-g}\nabla^m\nabla^n P_{amnb} \right); \qquad P_a^{\phantom{a}bcd}\equiv \left(\frac{\partial L}{\partial R^a_{\phantom{a}bcd} }\right)
\label{tpgrav3}
\end{equation}
and 
\begin{equation}
\delta v^j \equiv [2P^{ibjd}(\nabla_b\delta g_{di})-2\delta g_{di}(\nabla_cP^{ijcd})]
\label{genvc}
\end{equation}
The variation of the matter term will give 
\begin{equation}
\delta \mathcal{A}_m \equiv  -{1\over 2} \int T_{ik} \delta \gu ik \g d^4 x .
\label{qstressdef}
\end{equation}
where $T_{ik}$ is the symmetric stress tensor of matter defined by this relation.
To obtain the field equations from this variation, it is usual to assume that the boundary term involving $\delta v^j$ which arises in \eq{tpgrav2} can be ignored. 
In that case, the vanishing of the integrand of $\delta A_g$ requires the condition $[E_{ab} - (1/2)T_{ab}]\delta g^{ab} =0$. Since $\delta g^{ab}$ is symmetric, this leads to the result that the symmetric part  of $E_{ab}$ should be equal to $(1/2)T_{ab}$. 

In the Einstein's theory --- which is, of course, a special case of the above --- we have $E_{ab} = R_{ab} - (1/2) g_{ab} R$, which is known to be symmetric because Ricci tensor is symmetric. In the more general class of theories we are considering, it is not obvious from the expression in \eq{tpgrav3} that $E_{ab}$ is symmetric. A survey of the literature shows that this issue has not been clearly discussed and the field equation is written both in terms of $E_{ab}$ as well as in terms of the explicitly symmetrized form $E_{(ab)}$ by different authors. While carrying out the differentiation with respect to $g^{ab}$ in the first term of $E_{ab}$ in  \eq{tpgrav3}, one is led to the (symmetric) combination $\pc aijk R_{bijk} + \pc bijk R_{aijk}$. If the tensor  $\pc aijk R_{bijk}$ is symmetric in $a$ and $b$, then these two terms can be combined  into a single term, as is sometimes done in the literature (e.g., in my papers and textbook but without elaboration). However, it is not obvious why $\pc aijk R_{bijk}$ should be symmetric in $a$ and $b$  in general and I could not find an explicit proof in published literature.

The purpose of this note is to clarify many of these issues and prove several useful identities which help in the study of the generalized class of theories. These identities are non-trivial in the sense that they do not emerge from the \textit{algebraic} symmetry properties of $P_{abcd}$; instead they arise from the fact that when scalar quantities are constructed from several tensorial objects (like in the case of $L$ built from $R_{abcd}$ and $g^{ij}$), the derivatives of the scalar quantity with respect to these tensors must obey certain identities. While some of these results can be proved by other methods in special cases, the procedure I follow to derive these identities is very powerful and will be of use in more general contexts. (For example, the same procedure can be used  when the Lagrangian depends on the covariant derivatives of the curvature and I hope to discuss this in a future work.)

With this motivation, let us take a closer look at the variational principle and the resulting expressions mentioned above.

\subsection{Summary of results}

While varying $L\sqrt{-g}$ to obtain the equations of motion, we encounter two tensors closely related to the following   partial derivatives
\begin{equation}
 P^{abcd} = \pdlr{L}{R_{abcd}}_{g_{ij}}; \qquad P^{ab}=\pdlr{L}{g_{ab}}_{R_{ijkl}}
 \label{one}
\end{equation} 
The $P^{abcd}$ may be called the \textit{entropy tensor} of the theory because the integration of this tensor over  a horizon (contracted with a pair of binormals of the horizon) gives the entropy of the horizon in the generalized theories of gravity \cite{wald, tprop}. The derivatives appearing in \eq{one} are defined within the subspace of infinitesimal deformations which preserve the symmetries of the independent variable. Therefore,  $P^{ab}$ is symmetric while $P^{abcd}$ is assumed to inherit the algebraic symmetries of the curvature tensor: 
\begin{equation}
 P^{abcd} = -P^{bacd} = -P^{abdc};\quad  P^{abcd} = P^{cdab}; \quad  P^{a[bcd]} =0
 \label{symm}
\end{equation} 
One can construct several such tensors --- which have the properties listed in \eq{symm} --- from the curvature tensor and the metric. But these properties alone do \textit{not} guarantee that such a tensor can be expressed as the derivative of a scalar  with respect to the curvature tensor. The fact that $P^{abcd}$ has the form in \eq{one} leads to several \textit{further}  properties and inter-relationships which form the main thrust of this paper. In particular, we will prove the following results:

(1) Let us construct the tensor 
\begin{equation}
\mathcal{R}^{ab} \equiv P^{aijk} \rc bijk                                         
\end{equation} 
which plays the role of generalized Ricci tensor in these theories. We will prove that $P^{ab}=-2 \mr^{ab}$. That is, the derivatives of an \textit{arbitrary} scalar function with respect to the metric and the curvature are \textit{not} independent but are connected by the curious relation: 
\begin{equation}
 \pdlr{L}{g_{ab}}_{R_{ijkl}} = - 2 \rc bijk \pdlr{L}{R_{aijk}}_{g_{ab}}
 \label{key1}
\end{equation} 
 It immediately follows from \eq{key1} that the tensor $\mr^{ab}$ is symmetric. (This was noted earlier in a different context in \cite{koga}). This result does not follow merely from the algebraic symmetries
of $P^{abcd}$ given in \eq{symm} above.  The nature of the proof given in the next section indicates that it holds \textit{only because} $P^{abcd}$ is expressible as a derivative of a scalar with respect to the curvature tensor. 

(2) From \eq{key1} we can also obtain several other relations connecting the partial derivatives when different independent variables like 
 $(g^{ab}, R_{abcd})$ or $(g^{ab}, \rc abcd)$ or $(g^{ab},R^{ab}_{cd})$ are used to describe the system. 
 First, if we use the pair $(g^{ab}, R_{abcd})$ as independent variables, we have 
\begin{equation}
 \pdlr{L}{g^{im}}_{R_{abcd}} = - \pdlr{L}{g_{im}}_{R_{abcd}} = 2\mr_{im}
 \label{derdown}
\end{equation} 
Further, if we use the pair $(g^{ab}, \rc abcd)$ as independent variables, one can show 
\begin{equation}
 \pdlr{L}{g^{im}}_{\rc abcd} = \mr_{im}
 \label{deronethree}
\end{equation} 
and, more interestingly, if we use the pair $(g^{ab},R^{ab}_{cd})$ as independent variables, we get
\begin{equation}
 \pdlr{L}{g^{im}}_{R^{ab}_{cd}} = 0
 \label{dertwotwo}
\end{equation} 

An \textit{intuitive} way of understanding these results is as follows: To construct    scalar \textit{polynomials} of arbitrary degree in the curvature tensor,  using just the curvature tensor $R_{abcd}$ and metric $g^{ij}$ (with index placements as indicated), one requires two metric tensors to contract out the four indices of each curvature tensor in each term. 
So one can understand \eq{derdown} if $L$ was a polynomial in $R_{abcd}$.
An arbitrary scalar function of curvature and metric can be thought of having (possibly infinite) series expansion in powers of the curvature tensor. Since \eq{derdown} is linear in $L$, if it holds for an arbitrary polynomial in curvature tensor, then it should hold for arbitrary scalar functions which possess a series expansion. This intuitive argument finds strength in the result in \eq{dertwotwo} which says, for example, that in scalar polynomials constructed from $R^{ab}_{cd}$ the metric cannot appear and all the contractions must be with Kronecker delta! This is obvious from the fact that if we use $g^{ab}$ to contract any two lower indices, that will leave two upper indices hanging loose which cannot be contracted out because we do not have $g_{ab}$ available as an independent variable. In fact, if we assume \eq{dertwotwo} the other results can be obtained from it without  further ado.

(3) Using the fact that $P^{bcid}$ is anti-symmetric in $b$ and $c$, one can show 
\begin{equation}
 \nabla_b \nabla_c P^{bcid} 
 = \frac{1}{2} \left(\mr^{id} - \mr^{di}\right) = 0
 \label{eqnseven}
\end{equation} 
where the last equality follows from the symmetry of $\mr^{ab}$. Thus we get another counter-intuitive result that, for any scalar $L$ built from curvature tensor and the metric, we have the identity
\begin{equation}
 \nabla_a \nabla_b  \pdlr{L}{R_{abcd}}_{g_{ij}}=0
\end{equation} 
We shall now provide the proofs for these results. The  discussion in this paper will be based on the metric formalism in which the metric is treated as the fundamental variable in the Lagrangian. It is possible to approach these field theories in terms of affine-metric formalism or affine formalism wherein one deals with variational principles in which metric and the connection, say, are treated as independent (see e.g., \cite{affine}). I will not discuss these approaches though it should be possible to obtain similar results in these formalisms as well.

\section{Rigorous derivation of the results}

\subsection{The basic identities}
 
Consider an infinitesimal diffeomorphism $x^a\to x^a + \xi^a(x)$ which changes $L$, $g_{ab}$ and $R_{ijkl}$ by infinitesimal amounts.
The idea behind the proof of the identities is to express the Lie derivative of $L$ in two different  ways and equate the results. First, because $L$ is a scalar which depends on $x^i$ only through $g_{ab}(x)$ and $R_{ijkl}(x)$, if follows  that
\begin{equation}
\pounds_\xi L = \xi^m\nabla_m L = \xi^m P^{ab}\nabla_m g_{ab}+\xi^m P^{abcd} \nabla_m R_{abcd}
=P^{abcd} \xi^m\nabla_m R_{abcd}
\label{lie1}
\end{equation} 
where we have used $\nabla_m g_{ab}=0$.
On the other hand, 
if we think of the change $\delta L$ in $L$, due to small changes in $\delta g_{ab} \equiv \pounds_\xi g_{ab} $ and $\delta R_{ijkl} \equiv \pounds_\xi R_{ijkl}$,
we also know that 
\begin{equation}
 \pounds_\xi L = P^{ab}\pounds_\xi g_{ab} + P^{ijkl} \pounds_\xi R_{ijkl}
 \label{lie2}
\end{equation} 
We will now work out the right hand side of \eq{lie2} bringing it into a form similar to the right hand side of \eq{lie1} and equate the two expressions. That will lead to the result in \eq{key1}.

The first term in the right hand side of \eq{lie2}  is easy; using $ \pounds_\xi g_{ab}=\nabla_a\xi_b + \nabla_b \xi_a$ and the symmetry of $P^{ab} $ we get
\begin{equation}
 P^{ab}\pounds_\xi g_{ab} = 2 P^{ab} \nabla_a \xi_b
 \label{plg}
\end{equation} 
The term involving Lie derivative of curvature tensor requires more work. The Lie derivative of the curvature tensor will have one term with the structure $\xi^m \nabla_m R_{ijkl}$ and four other terms having the structure $R \nabla \xi$. On contracting with $P^{ijkl}$ and using \eq{symm} we can combine these four terms pairwise and obtain 
\begin{equation}
 P^{ijkl}\pounds_\xi R_{ijkl} = P^{ijkl}\xi^m \nabla_m R_{ijkl} + 2P^{ijkl}\left[ (\nabla_i \xi^m) R_{mjkl} + (\nabla_k \xi^m) R_{ijml}\right]
 \label{lie3}
\end{equation} 
Relabeling the dummy indices on $\nabla\xi$ and using the symmetries $R_{mlij} = R_{ijml}$ and $P^{kjil}= P^{ilkj}$ we get
\begin{equation}
 P^{ijkl}\pounds_\xi R_{ijkl} = P^{ijkl}\xi^m \nabla_m R_{ijkl}+ 4  (\nabla_i \xi_m)\mr^{im}
 \label{plr}
\end{equation} 
where we have used the definition $\mr^{im} = P^{ijkl} \rc mjkl$.
Using \eq{plr} and \eq{plg} in \eq{lie2} we get 
\begin{eqnarray}
 \pounds_\xi L &=& P^{ijkl} \xi^m \nabla_m R_{ijkl} + 2 (\nabla_i \xi_m) [P^{im} + 2 \mr^{im}]\nonumber\\
 &=& \pounds_\xi L + 2 (\nabla_i \xi_m) [P^{im} + 2 \mr^{im}]
\end{eqnarray} 
Since $\xi_m$ is arbitrary, it follows that the term within square brackets in the second line has to vanish,  which leads to  the condition  in \eq{key1}:
\begin{equation}
 P^{im}= \pdlr{L}{g_{im}}_{R_{abcd}} = -2 P^{ijkl} \rc mjkl = - 2 \mr^{im}
 \label{key3}
\end{equation} 
Since the left hand side is symmetric, it  follows that $\mr^{im}$ is symmetric. 

Equation (\ref{eqnseven}) can be obtained immediately from the above result. To do this, we start with the relation:
\begin{eqnarray}
 \nabla_b\nabla_c P^{bcid} &=& \frac{1}{2} \big[ \nabla_b,\nabla_c\big] P^{bcid}\nonumber\\
 &=& \frac{1}{2} \bigg\{ \rc bkbc P^{kcid} + \rc ckbc P^{bkid} + \rc ikbc P^{bckd} + \rc dkbc P^{bcik}\bigg\}
\end{eqnarray} 
The first equality uses $P^{bcid} = - P^{cbid}$ and second is a standard result for commutator of covariant derivatives.
The first two terms vanish since $\rc bkbc = - \rc bkcb = R_{kc}$ is symmetric while $P^{ijmn} = - P^{jimn}$. Hence 
\begin{eqnarray}
 \nabla_b\nabla_c P^{bcid} &=&   \frac{1}{2}\bigg\{\rc ikbc P^{kdbc} + \rc dkbc P^{ikbc}\bigg\}
 =\frac{1}{2}\bigg\{ - \rc ikbc P^{dkbc} +\rc dkbc P^{ikbc}\bigg\}\nonumber\\
 &=& \frac{1}{2} \Big(- \mr^{di} + \mr^{id}\Big) =0
 \label{twenty}
\end{eqnarray}
To obtain the first equality we have used the pair exchange symmetry of $P^{ijkl}$ in both terms and to obtain the second equality we have used $P^{kdbc}=- P^{dkbc}$.

From \eq{key3}, one can derive a few corollaries which are useful in expressing the field equations that arise from our Lagrangian.
We first note that since $\delta g_{ab} = - g_{ai} g_{bj} \delta g^{ij}$ we have: 
\begin{equation}
  \pdlr{L}{g^{im}}_{R_{abcd}} =- \pdlr{L}{g_{im}}_{R_{abcd}} =2 P_i^{\phantom{i}jkl}R_{mjkl}= 2\mr_{im}
\end{equation} 
We next consider how our results change if we use the pair ($g^{im}, \rc abcd$) or the pair $(g^{ab}, R^{ij}_{kl})$ as the independent  variables.
It is obvious that $\pdsl{L}{g^{im}} $ with $R_{abcd} $ held fixed is \textit{not} the same as 
$\pdsl{L}{g^{im}} $ with $\rc abcd$ held fixed. To find their inter-relationship, we note that 
\begin{eqnarray}
 dL &=& \pdlr{L}{g^{im}}_{R_{abcd}} dg^{im} + P^{abcd}  dR_{abcd} = \pdlr{L}{g^{im}}_{\rc abcd} dg^{im} + P_a^{\phantom{a}bcd} d(g^{ak} R_{kbcd})\nonumber\\
 &=& P^{mbcd} dR_{mbcd} + dg^{im} \left\{ \pdlr{L}{g^{im}}_{\rc abcd} + \mr_{im} \right\}
 \label{interrelation}
\end{eqnarray}
It follows that
\begin{eqnarray}
  \pdlr{L}{g^{im}}_{\rc abcd}  &=& \pdlr{L}{g^{im}}_{R_{abcd}} - \mr_{im} \nonumber\\
  &=& 2 \mr_{im} - \mr_{im} = \mr_{im} = P_i^{\phantom{i}bcd}R_{mbcd} 
  \label{five}
\end{eqnarray}
More interestingly, if we work with $(g^{ab}, R^{ij}_{kl})$ as the independent  variables we get in place of \eq{interrelation} the relation:
\begin{equation}
 dL = \pdlr{L}{g^{im}}_{R_{abcd}} dg^{im} + P^{abcd} dR_{abcd} = P^{abcd} dR_{abcd} + dg^{im} \left\{\pdlr{L}{g^{im}}_{R^{ab}_{cd}} + 2 \mr_{im}\right\}
\end{equation} 
leading to
\begin{equation}
 \pdlr{L}{g^{im}}_{R^{ab}_{cd}} = \pdlr{L}{g^{im}}_{R_{abcd}} - 2 \mr_{im}=0
\end{equation} 
As mentioned earlier, this result shows that scalars built from $g^{ij}, R^{ab}_{cd}$ are actually independent of the metric tensor when $R^{ab}_{cd}$ is held fixed.

For the sake of completeness, I should mention the following subtlety as regards the very first equation \eq{lie1} which, I believe, the reader thought was obvious; but it requires some pedagogical discussion. Consider a scalar function $f(u^i, A_{ij}, B_i^{\phantom{i}jk}, ....)$ built out of several tensorial quantities $u^i(x), A_{ij}(x), B_i^{\phantom{i}jk}(x), ... $. We assume that the $f$ depends on $x^i$ only implicitly through these tensorial objects. It then follows from standard laws of calculus that the \textit{partial} derivative
$\partial_i f$ can be expressed in terms of the partial derivatives of $f $ with respect to $u^i, A_{ij}, B_i^{\phantom{i}jk} ....$ and the partial derivatives of the tensorial independent variables. Can one write a similar relation using \textit{covariant} derivatives rather than partial derivatives? It turns out that one indeed can though the result is not trivial because the simultaneous validity of the two results --- one involving ordinary derivatives and one involving covariant derivatives --- again leads to certain identities amongst the derivatives of $f$ with respect to the tensorial independent variables. The validity of the `function of the function' rule for covariant differentiation --- which again I have not seen discussed explicitly in text books --- can be proved in two ways: 
(a) In the local inertial frame, the relation with covariant derivatives reduces to the one with partial derivatives, the validity of which is obvious from ordinary laws of calculus. The general covariance of the expression assures the validity of the result in any coordinate system. 
(b) A more explicit proof which explains the ``mechanism'' behind the result begins by noticing that to construct a scalar out of tensorial quantities one needs to first construct a set of scalars involving \textit{products} of the tensorial quantities by contracting on all indices like, for example, $\phi_1 = u^i u^j A_{ij}, \ \phi_2 =  B_i^{\phantom{i}jk} A_{jk}u^i ...., $ and then construct our scalar as a function $f(\phi_1, \phi_2, ....)$ of primitive scalars which only involve products of tensorial quantities. We know, of course, that any derivative (partial, covariant, Lie) obeys the \textit{product} rule. Given the fact that any scalar $f$ has to be a function of primitive scalars $(\phi_1, \phi_2, ....)$ which must be \textit{products} of the tensorial quantities, we obtain the function of the function differentiation rule for covariant derivatives of scalars from the product rule of covariant differentiation. Once again we see the importance of combinatorics ensuring that all the indices must be contracted in the construction of scalars.

\subsection{Field equations in the generalized theory of gravity}

To obtain the field equation by varying the Lagrangian, the most convenient set of variables to use happens to be the pair ($g^{im}, \rc abcd$). This is because $\rc abcd$ can be expressed entirely in terms of $\Gamma^i_{jk}$ without the use of a metric which makes the variations simple to perform. 
As mentioned earlier \cite{TPtext}, the equations of motion obtained by varying $L\g$ can be expressed in the form $E_{ab} = (1/2) T_{ab}$ where $T_{ab}$ is the energy-momentum tensor of matter and
\begin{eqnarray}
 \g E_{ab} &=& \pdlr{L\g}{g^{ab}}_{\rc ijkl} - 2 \g \nabla^m \nabla^n P_{amnb}\nonumber\\
 &=& \g \left[ \pdlr{L}{g^{ab}}_{\rc ijkl} - \frac{1}{2} g_{ab} L - 2 \nabla^m \nabla^n P_{amnb}\right]
\end{eqnarray}
Using \eq{five} we can express this in a simple form as 
\begin{equation}
 E_{ab} 
 = \mr_{ab} - \frac{1}{2} g_{ab} L -  2 \nabla^m \nabla^n P_{amnb}
  \label{seven}
\end{equation} 
which does not involve the derivatives of the Lagrangian. This is the form which I believe is most useful for computational purposes, though in the literature one often finds more complicated expressions possibly because the issues addressed above were not clearly understood.

The equation $E_{ab} = (1/2) T_{ab}$ with a symmetric energy momentum tensor requires  $E_{ab}$ to be symmetric and have already proved that $\mr_{ab}$ is symmetric. This implies that the last term $\nabla_m \nabla_n P^{amnb}$ in \eq{seven} is also symmetric in $a$ and $b$. It is easy to see that the anti-symmetric part of this tensor is given by 
\begin{equation}
 \nabla_m \nabla_n \left( P^{amnb} - P^{bmna}\right) =\nabla_m \nabla_n \left( P^{amnb} + P^{anbm}\right) 
 = -\nabla_m \nabla_n P^{abmn}
  \label{eight}
\end{equation} 
In arriving at the first equality we have used the pair exchange symmetry and the anti-symmetry of the first two indices $P^{ijkl}$; in arriving at the second equality we have used the cyclic relation 
$P^{i[jkl]} =0$. It follows that the last term in \eq{seven} is symmetric only if the right hand side of \eq{eight} vanishes which is assured by \eq{twenty} and the symmetry of $P^{abcd}$ under pair exchange. Thus each of the terms in \eq{seven} is individually symmetric.

From the expression \eq{tpgrav2} for the variation of the action, one can obtain a generalized Bianchi identity satisfied by $E^a_b$. If we consider the variation $\delta g^{ab} = \nabla^a \xi^b +\nabla^b \xi^a  $ arising from the coordinate change $x^a \to x^a+\xi^a$ and assume that $\xi^a$ and its derivatives vanish sufficiently fast close to the boundary of the region, we get the result 
\begin{equation}
 \nabla_a E^a_b = \nabla_a \left[\mathcal{R}^a_b - \frac{1}{2}\delta^a_b  L -  2 \nabla^m \nabla^n P^a_{\phantom{a}mnb}\right] =0
\end{equation} 
Once again it is not obvious that the expression in the square bracket is divergence-free, though it is. Explicit covariant differentiation of the expression will again lead to derivatives of $L$ and the generalized Bianchi identity will be satisfied for reasons similar to the one described above.

The first two terms in \eq{seven} for $E_{ab}$ involve at most second derivatives of the metric tensor with respect to the coordinates while the third term $\nabla\nabla P$ could involve up to fourth derivative of the metric tensor with respect to the coordinates. A very important subclass of generalized theories of gravity is obtained by ensuring that the field equations do \textit{not} contain more than second order derivatives. Obviously, such a class of theories can be obtained from the Lagrangian which satisfy the additional constraint $\nabla_a P^{abcd}=0$.  Given the symmetries of $P^{abcd}$, it follows that ($\partial L /\partial R_{abcd}$) must be a divergence-free tensor in all indices. In such a case, the field equations simplify significantly and we get 
\begin{equation}
 E_{ab} 
 = \mr_{ab} - \frac{1}{2} g_{ab} L 
 = \pc aijk R_{bijk} - \frac{1}{2} g_{ab} L 
 = \frac{1}{2} T_{ab}
  \label{sevenone}
\end{equation} 
The similarity with Einstein's theory is obvious. In this context we have constructed from $L$ a second rank  tensor $E^{ab}$ which (a) is symmetric, (b) divergence-free and (c) does not contain more than second derivatives of the metric. It can be proved \cite{lovelock} that the most general tensor which satisfies these criteria is the \LL\ tensor obtained from the \LL\ Lagrangian. Therefore, the theory in which $P^{abcd}$ is divergence-free is the same as \LL\ models of gravity and the corresponding $L$ must be the \LL\ Lagrangian.

All these theories possess a Noether current which is conserved off-shell due to the diffeomorphism invariance of the Lagrangian. The identities obtained above have important implications for the structure of Noether current and for derivation of field equations in the emergent paradigm \cite{tprop,tp1,tp2} starting from the Noether current. This will be examined in a separate publication.

\section{Some further  curiosities  and an alternative derivation} 

While the analysis given above seems both algebraically and logically straightforward, there are some subtleties about these results which we will now discuss. First, one would have thought that some physical meaning or intuitive content could be attributed to the result in \eq{twenty} which looks strikingly similar to some kind of Bianchi identity.  Unfortunately, I could not find any simple interpretation of this very general result. Obviously, the result relies on the symmetries of $\mr^{ab}$ which, in turn, arises from the relationship in \eq{key1} and does not have a simple intuitive interpretation. This aspect deserves further examination. (This result is, of course, related to the  generalized Bianchi identity, $\nabla_a E^{ab}=0$ but I could not make the connection self-evident.)

The second subtlety about these results is the occurrence of the factor $\nabla_i \xi_m$ in \eq{plr} which allows us to obtain \eq{key1}. This is surprising for the following reason: We know that, under the  diffeomorphism $x^i \to x^i + \xi^i$, the metric changes by 
\begin{equation}
\delta g_{ab} = -[\nabla_a \xi_b + \nabla_b \xi_a] = - \pounds_\xi g_{ab}                                                                             \end{equation} 
showing that  $\delta g_{ab}$ and $\pounds_\xi g_{ab}$ depend only on the symmetric part of $\nabla_a\xi_b$. It immediately follows that $\delta \Gamma^i_{jk}$ as well as $\delta R_{abcd}$
can depend \textit{only} on the symmetric part of $\nabla_i \xi_m$. Therefore, the left hand side of \eq{plr} can depend only on the symmetric part of $\nabla_i \xi_m$. 

 Note that if we had only the symmetric part of  $\nabla_i \xi_m$ in the second term of \eq{plr} --- rather than $\nabla_i \xi_m $ itself --- we could not have obtained the result in \eq{key3}; instead we would have found that the symmetric part of $\mr^{im}$ is related to $P^{im}$ and one cannot prove the symmetry of $\mr^{im}$.
It is only the separation of the left hand side of \eq{plr} into two terms on the right hand side with a very specific structure for the first term, which introduces $\nabla_i \xi_m$ in \eq{plr}. In other words, it is the first term in the right hand side of \eq{plr} which breaks the inherent symmetry and allows us to obtain the result which, of course, is closely related to the fact that $L$ is a scalar.
So, this subtlety, fortunately finds a resolution in the above algebraic fact. 

It is nevertheless worth checking this fact explicitly by evaluating $\pounds_\xi R_{abcd}$ in terms of $\pounds_\xi g_{ab}$ and rewriting the expression to obtain \eq{plr}. Since the variations in the curvature tensor and the metric are related in the form $\pounds_\xi R = \nabla \nabla (\pounds_\xi g)$  we would expect $P \pounds_\xi R = P \nabla \nabla \nabla \xi$, which can depend only on the symmetric part of $\nabla \xi$ (when we re-introduce indices correctly!). \textit{But 
at this stage it has no $P\xi \nabla R$ type of terms which is the first term in the right hand side of \eq{plr}.} When we commute the $\nabla$s properly and use the identities obeyed by the curvature tensor,
 one can rewrite this expression with a $P\xi \nabla R$ term separated out. But then, the remaining terms depend on full $\nabla \xi$ rather than on the symmetric part alone!  

This is to be expected because the two ways of computing the Lie derivative of curvature tensor must give the same result. The first way is to treat it as a fourth rank tensor which will indeed introduce a $\xi \nabla R $ term in the Lie derivative almost by definition. The second way is to express $R$ in terms of the metric and compute the result which has to come from an expression with the structure $\nabla \nabla \nabla \xi$. Either way, the result for $\pounds_\xi R_{abcd}$ can only depend on the symmetric part of $\nabla \xi$. But if we subtract out the $\xi \nabla R $ term, the remaining terms, of course, can depend on the full $\nabla \xi$ rather than on its symmetric part. This is what happens.

We will now outline the steps in this derivation for the sake of completeness. It is easier to start with the expression for $P_a^{\phantom{a}bcd}\delta \rc abcd $ rather than from $P^{abcd}\delta R_{abcd}$. (These two are, of course, \textit{ not} the same when the metric is varied; see \eq{thirty} below). We begin with the standard result (see, e.g.,\cite{TPtext})
\begin{equation}
 -P_a^{\phantom{a}bcd} \delta \rc abcd = - 2 P^{ibcd} \nabla_c \nabla_b \delta g_{di} = \left( 2  P^{ibcd} \nabla_c \nabla_b \nabla_d \xi_i + 2P^{ibcd} \nabla_c \nabla_b \nabla_i \xi_d\right)
\end{equation} 
In the triple $\nabla$s  we want to commute the indices so that the anti-symmetry of $P^{ijkl}$ can be utilized. Doing this and using  $[\nabla_i, \nabla_j] \xi^l=\rc lkij \xi^k $ we get
\begin{eqnarray}
 -P_a^{\phantom{a}bcd} \delta \rc abcd &=&  2 P^{ibcd} \nabla_c \Big\{ \nabla_d\nabla_b + \left[ \nabla_b, \nabla_d\right] \Big\} \xi_i + P^{ibcd} \nabla_c \Big\{  \left[ \nabla_b, \nabla_i\right]\xi_d\Big\}\nonumber\\
 &=& P^{ibcd} \left[ \nabla_c, \nabla_d\right] (\nabla_b \xi_i)  - P^{ibcd} \nabla_c \Big\{ (2 \rc mibd + \rc mdbi)\xi_m \Big\}
\end{eqnarray} 
The second term in the right hand side can be replaced by $- P^{ibcd} \nabla_c [2\rc mdbi \xi_m]$. To see this, we use the anti-symmetry of $P^{ibcd}$ to write $2 \rc mibd $ as the sum of $\rc mibd$ and $-\rc mbid = \rc mbdi$ and use the cyclic property of the curvature tensor. This gives 
\begin{eqnarray}
 P^{ibcd} \nabla_c \Big[ \left( 2 \rc mibd  + \rc mdbi \right) \xi_m \Big] 
 &=& P^{ibcd} \nabla_c \Big[ \left(  \rc mibd  +\rc mbdi + \rc mdbi \right) \xi_m \Big] \nonumber\\
 &=& P^{ibcd} \nabla_c \Big[ \left( - \rc mdib  +  \rc mdbi \right) \xi_m \Big] \nonumber\\
 &=& P^{ibcd} \nabla_c \left( 2 \rc mdbi \xi_m\right)
 \label{secterm}
\end{eqnarray}
This will expand to a structure like $P \xi \nabla R + P R \nabla \xi$. In the term involving $\xi\nabla R $ we want to bring in the directional derivative $\xi^m \nabla_m$. This can be done using the result 
\begin{equation}
 2 P^{ibcd} \xi_m \nabla_c \rc mdbi = - P^{ibcd} \xi^m \nabla_m R_{ibcd}
 \label{dirderive}
\end{equation} 
The proof of \eq{dirderive} is similar to that for \eq{secterm}. We have, 
\begin{eqnarray}
 2P^{ibcd} \nabla_c R_{mdbi} &=& P^{ibcd} \left( \nabla_c R_{mdbi} - \nabla_d R_{mcbi}\right) 
 = P^{ibcd} \left( \nabla_c R_{bimd} - \nabla_d R_{bimc}\right) \nonumber\\
 &=& - P^{ibcd} \nabla_m R_{ibcd}
\end{eqnarray}
In arriving at the last step we have used 
\begin{equation}
 -\nabla_d R_{bimc} = \nabla_c R_{bidm} + \nabla_m R_{bicd}
\end{equation} 
and cancelled out one term using the anti-symmetry in $d$ and $m$. Plugging all these back, we get 
\begin{eqnarray}
 -P_a^{\phantom{a}bcd} \delta \rc abcd &=& P^{ibcd} \xi^m \nabla_m R_{ibcd}  - P^{ibcd} \left(\rc mbcd \nabla_m \xi_i + \rc micd \nabla_b\xi_m + 2 \rc mdbi \nabla_c \xi_m \right)\nonumber\\
 &=& P^{ibcd} \xi^m \nabla_m R_{ibcd} - (\nabla_r \xi_s) \Big[ P^{sbcd} \rc rbcd + \rc sicd P^{ircd} + 2 \rc sdbi P^{ibrd}\Big]\nonumber\\
\end{eqnarray}
The first term is in the required form and the second term does depend on $\nabla_m \xi_i$ rather than on its symmetric part alone. We will now simplify this term a little bit \textit{without}, of course, assuming that $P^{ijkl} \rc mjkl $ is symmetric in $i$ and $m$ since we intend this to be an independent derivation of \eq{key3}. In the second term on the right hand side, we write $P^{ircd} = - P^{ricd}$ and $P^{ibrd} = - P^{rdbi}$. This allows us to combine two of the terms getting the final expression to be
\begin{equation}
 -P_a^{\phantom{a}bcd} \delta \rc abcd =  P^{ibcd} \xi^m \nabla_m R_{ibcd} - (\nabla_r \xi_s) \Big[ P^{sbcd} \rc rbcd -3  \rc sbcd P^{rbcd}\Big]
 \label{fourteen}
\end{equation} 
In spite of appearances, we know that the right hand side can  depend only on the symmetric part of $\nabla_r \xi_s$ but the individual terms separately do not obey this restriction. The result is very similar to the one in \eq{plr} but not exactly the same because the left hand sides are slightly different. To make the final step, we note that 
\begin{equation}
 P^{ijkl} \delta R_{ijkl} = P^{ijkl} \delta ( g_{lm} \rc mjkl ) = P_m^{\phantom{m}jkl} \delta \rc mjkl - P^{ijkl} \rc mjkl (\nabla_i \xi_m + \nabla_m \xi_i)
 \label{thirty}
\end{equation}
Substituting for the first term on the right hand side of \eq{thirty} from  \eq{fourteen}, we get 
\begin{eqnarray}
 P^{ijkl} \delta R_{ijkl}& =& -P^{abcd} \xi^m \nabla_m R_{abcd} + (\nabla_r\xi_s) \left[ \rc rbcd P^{sbcd} - 3 \rc sbcd P^{rbcd}\right]\nonumber\\
 && \qquad\quad - P^{ijkl} \rc mjkl (\nabla_i \xi_m + \nabla_m \xi_i)
\end{eqnarray} 
When we change the dummy indices $i$ and $m$ to $r$ and $s$ properly in the last term and combine it with the second term, we find that two terms add up while another pair of terms cancel (all without assuming symmetry of $P^{sbcd} \rc rbcd$ in $r, s$). Thus we again get
\begin{equation}
 P^{ijkl} \delta R_{ijkl}= - P^{ijkl} \pounds_\xi R_{ijkl} = - P^{abcd} \xi^m \nabla_m R_{abcd} - 4 (\nabla_r \xi_s) \rc sbcd P^{rbcd} 
 \label{seventeen}
\end{equation} 
which matches with \eq{plr}.
Once again, from the very nature of the derivation, we know that the right hand side depends only on the symmetric part of $\nabla_r \xi_s$ but the individual terms do not maintain this symmetry. This result, of course, is identical to the one obtained in \eq{lie3}. In the original derivation of \eq{key3} it was not obvious that the right hand side can depend only on the symmetric part of $\nabla\xi$. But the algebra was simpler. The derivation of \eq{seventeen} is more involved but makes clear that the right hand side can only depend on the symmetric part of $\nabla \xi$.

In deriving either of these, \eq{key3} or \eq{seventeen}, we have only used the symmetries of $P^{abcd}$ but not the fact that it is the derivative of a scalar with respect to the curvature tensor. It is this vital input (plus the facts that $L$ has no explicit dependence on $x^i$ and $\nabla g =0$) which allows us to separate out the first term and obtain the condition in \eq{key3}.

\section*{Acknowledgments}

I thank D. Kothawala, D. Lynden-Bell, A. Paranjape for useful and extensive discussions on these results and Hamsa Padmanabhan and Krishna Parattu for comments on an earlier draft. This work was completed while I was visiting Institute of Astronomy, Cambridge during May-July, 2011 and the hospitality of IOA, Cambridge is gratefully acknowledged. My research is partially supported by J.C. Bose Fellowship of DST, India.


\begin{thebibliography}{99}

\bibitem{tprop} T. Padmanabhan, \textit{Rep. Prog. Phys.}, \textbf{73}, 046901 (2010) [arXiv:0911.5004]; T. Padmanabhan , \textit{Lessons from Classical Gravity about the Quantum Structure of Spacetime}, \textit{J.Phys. Conf.Ser.}  \textbf{306}  012001 (2011) [arXiv:1012.4476]. 

\bibitem{TPtext} For a textbook derivation, see page 657 of T. Padmanabhan, \textit{Gravitation: Foundations and Frontiers}, (Cambridge University Press, 2010).

\bibitem{affine}
G. J. Olmo, 2011, \textit{Int.J.Mod.Phys}, \textbf{D20}, 413 [arXiv:1101.3864]; 
S. Capozziello, M. De Laurentis, 2011, \textit{Extended Theories of Gravity}, [arXiv:1108.6266].
 
\bibitem{wald} 
Wald R M 1993 \textit{Phys. Rev. D}   {\bf  48}  3427 [gr-qc/9307038];
Iyer V  and  Wald R M 1995 \textit{Phys. Rev. D} {\bf 52}  4430 [gr-qc/9503052].

\bibitem{koga}
J. Koga and K. Maeda,  \textit{Phys. Rev.} \textbf{D 58}, 064020 (1998)

\bibitem{tp1} T. Padmanabhan, \textit{A Physical Interpretation of Gravitational Field Equations}, AIP Conference Proceedings,  \textbf{1241},  93-108 (2010) [arXiv:0911.1403] 

\bibitem{tp2} T. Padmanabhan,    \textit{Phys.Rev.}, \textbf{D 81},, 124040 (2010) [arXiv:1003.5665]  

 \bibitem{lovelock} 
Lanczos C   {\it Z. Phys.}, {\bf 73}, 147 (1932); 
Lovelock D   {\it J. Math. Phys.} {\bf 12} 498 (1971); 
D. Lovelock and H.Rund \textit{Tensors, Differential forms and Variational Principles} (Dover, 1989).


\end{thebibliography}
\end{document}